\documentclass[reprint,superscriptaddress,amsmath,amssymb,aps,prl]{revtex4-2}
\usepackage{graphicx}
\usepackage{dcolumn}
\usepackage{bm}
\usepackage{color}
\usepackage{units}
\usepackage{wasysym}
\usepackage{MnSymbol}
\usepackage{comment}
\usepackage{times}
\usepackage{amsmath}
\usepackage{mathtools}
\usepackage[unicode=true,pdfusetitle,bookmarks=false,colorlinks=true,citecolor=blue,urlcolor=blue,linkcolor=blue]{hyperref}
\begin{document}

\title{Topological Forces in a Model System for Reptation Dynamics}
\author{Ahmad K. Omar}
\email{aomar@berkeley.edu}
\affiliation{Department of Materials Science and Engineering, University of California, Berkeley, California 94720, USA}
\affiliation{Materials Sciences Division, Lawrence Berkeley National Laboratory, Berkeley, California 94720, USA}
\author{Yuyuan Lu}
\email{yylu@ciac.ac.cn}
\affiliation {State Key Laboratory of Polymer Physics and Chemistry, Changchun Institute of Applied Chemistry, Chinese Academy of Sciences, Changchun 130022, China}
\author{Lijia An}
\email{ljan@ciac.ac.cn}
\affiliation {State Key Laboratory of Polymer Physics and Chemistry, Changchun Institute of Applied Chemistry, Chinese Academy of Sciences, Changchun 130022, China}
\author{Zhen-Gang Wang}
\email{zgw@caltech.edu}
\affiliation {Division of Chemistry and Chemical Engineering, California Institute of Technology, Pasadena, California 91125, USA}
\affiliation {State Key Laboratory of Polymer Physics and Chemistry, Changchun Institute of Applied Chemistry, Chinese Academy of Sciences, Changchun 130022, China}

\begin{abstract}
We construct a micromechanical version of an early model for topologically constrained polymers -- a 2D chain amongst point-like uncrossable obstacles -- which allows us to explicitly elucidate the role of topological forces beyond confining the chain to a curvilinear tube-like path. 
Our simulations reveal that linear relaxation of the contour length \textit{along the tube} is slowed down by the presence of topological forces that can be considered as additional effective topological ``friction'' in quiescence.
However, this perspective fails in predicting the strong forces that resist the imposed curvilinear motion of the chain during nonlinear startup microrheology. 
These entropic forces are nonlocal in nature and result from an unexpected coupling between orientational and longitudinal dynamics.
\end{abstract}
\maketitle

\section{Introduction and Motivation}

Understanding the motion of macromolecules in topologically restrictive environments is a problem of fundamental and practical interest and has been an ongoing challenge for over a half century. 
In a concentrated solution or melt consisting of sufficiently long polymer chains, the motion of a test chain becomes restricted at a length scale $d_T$ (the tube diameter) due to interchain uncrossability, which forces the test chain to diffuse along a curvilinear tube-like path (the primitive path). 
The motion of a test chain along the tube is presumed to proceed via the same dynamics that govern the motion of the chain on length scales below $d_T$~\cite{Doi1986,Rubinstein2003}. 
This molecular picture for the motion of a topologically constrained polymer was provided by de Gennes, Doi and Edwards~\cite{Gennes1971, Doi1986} and motivated the highly successful~\cite{Watanabe1999,McLeish2002} phenomenological Doi-Edwards (DE) tube model~\cite{DoiEdwards1978, *DoiEdwards1978a, *DoiEdwards1978b, *DoiEdwards1979} (and its modern adaptations~\cite{Milner1998, Graham2003}) for entangled polymer dynamics and rheology.

Despite the success of the constitutive equations developed using the above physical picture, little is known about the nature of the very topological forces that give rise to the confining tube. 
For instance, this perspective implicitly assumes an infinite (harmonic) confining tube potential acting in the transverse direction of the primitive path and the absence of topological forces entirely in the longitudinal direction (resulting in ``barrier-free" curvilinear dynamics). 
These underlying assumptions regarding topological forces have been revisited in the last decade, with Sussman and Schweizer self-consistently determining the transverse tube potential for entangled polymers to be strongly anharmonic via a dynamical mean-field theory~\cite{Sussman2012a, Sussman2013}, in agreement with simulation~\cite{Tzoumanekas2006,Zhou2006}. 
While some have speculated that there may be a longitudinal ``friction" imposed by the topological constraints that can quantitatively modify curvilinear relaxation~\cite{Likhtman2007, QinMilner2016}, such friction has yet to be quantified and remains a conjecture. 
The role of topological forces in systems far from equilibrium remains an open question.
Further, some observations made in nonlinear rheological experiments~\cite{Wang2017, Boukany2009} and simulations~\cite{Xu2018} of entangled polymers can be difficult to reconcile with existing tube physics. 
This has led some~\cite{Wang2007,Wang2015, Schweizer2018} to speculate that fast deformation can generate strong topological forces (a ``grip force") that induce chain stretching and inhibit stretch relaxation; however this idea has yet to be verified by simulation. 

Clearly, resolving if topological forces play a role -- beyond confining the chain to the primitive path -- in both linear and nonlinear polymer dynamics remains an outstanding challenge. 
In this Article, we aim to provide quantitative insight into this problem by returning to the original model for reptation physics first examined by de Gennes~\cite{Gennes1971} and later used to confirm several of its predictions~\cite{EvansEdwards1981, EdwardsEvans1981, EvansEdwards1981a, Deutsch1989a, Reiter1991, Ianniruberto1998, Shie2010} and explore a variety of phenomena~\cite{OlveradelaCruz1986,Baumgaertner1987,Hickey2007}: a linear chain moving in a 2D array of uncrossable point obstacles [see Fig.~\ref{fig:pcanalysis} or Fig.~\ref{fig:relaxation}(a)]. 
We construct a micromechanical version of this model to explicitly compute the topological \textit{forces} acting on the chain. 
We find that relaxation of the chain along the primitive path is quantitatively slowed in comparison to the expected Rouse relaxation.  
We interpret this finding as evidence of a curvilinear \textit{topological friction}.
Nonlinear microrheology, however, reveals that this ``friction" is inherently nonlocal and can generate an entirely non-Rouse response far from equilibrium. 

\section{A Model with Explicit Topolgical Forces}

\subsection{Micromechanical Model}

We consider a linear chain of $N$ connected ideal (i.~e., volume-less) ``beads" in the presence of uncrossable point obstacles arranged in an infinite square lattice with a lattice constant $d_T$  [see Fig.~\ref{fig:pcanalysis} or Fig.~\ref{fig:relaxation}(a)]. 
The bead dynamics follow the overdamped Langevin equation
\begin{equation}
	\label{eq:EOM}
	\zeta_0\bm{\dot{x}} =\bm{f_b}+\bm{f_c}+\bm{f_o}, 
\end{equation}
where $\bm{\dot{x}}$ is the bead velocity, $\zeta_0$ is the drag coefficient while $\bm{f_b}$, $\bm{f_c}$ and $\bm{f_o}$ represent forces from Brownian fluctuations (taken to be a white noise with a mean of $\bm{0}$ and variance of $2kT\zeta_0\mathbf{I}$ where $kT$ is the thermal energy and $\mathbf{I}$ is the identity tensor), chain connectivity and obstacle uncrossability, respectively.
Beads are connected with a stiff harmonic spring with spring constant of $30kT/\ell^2$ (where $\ell$ is the spring rest length), resulting in an average bond (Kuhn) length of $b \approx 1.01\ell$. 
We choose $\ell$, $kT$, $\zeta_0 \ell^2 /kT$, respectively, as the units of length, energy and time. 

\begin{figure}
	\centering
	\includegraphics[width=0.48\textwidth,keepaspectratio,clip]{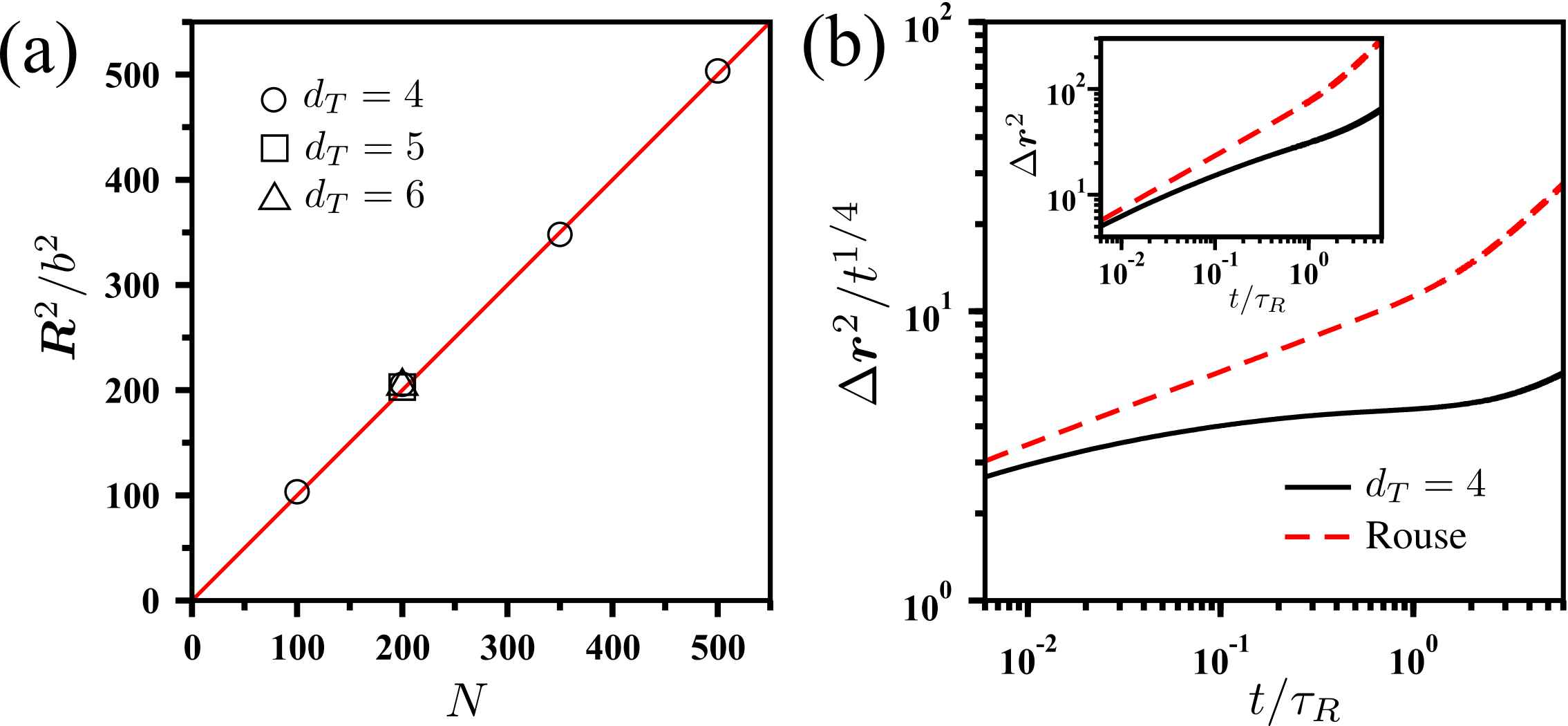}
	\caption{{\protect\small{(a) Chain conformation in the presence of obstacles. The line shown has a slope of one. (b) MSD of the interior $N_e = 16$ segments normalized by the expected $t^{1/4}$ scaling for $N=200$ with ($d_T = 4$) and without (Rouse) obstacles. The unnormalized MSD is shown in the inset.}}}
	\label{fig:diffusion}
\end{figure}

To enforce uncrossability while maintaining the point-like nature of the obstacles, we use a procedure that is similar to the potential-free algorithm for simulating hard-spheres~\cite{Heyes1993, Foss2000}. 
We first integrate Eq.~\eqref{eq:EOM} for each bead over a time step $\Delta t$ without considering obstacles. 
If we detect that a bond has crossed an obstacle, we return the two beads making up the bond to their pre-integrated locations. 
The force exerted by the obstacle on a bead $\bm{f_o}$ is then simply the force required to return the bead to its original location ($\bm{f_o} = -\zeta_0\Delta\bm{x}/\Delta t$ where $\Delta\bm{x}$ is the original bead displacement). 
In addition to allowing for the explicit determination of topological forces, this method ensures that all forces in our system are \textit{entropic}, arising from either chain connectivity or topology. 

The point-like nature of the obstacles should not alter the equilibrium distribution of chain conformations while altering the dynamical landscape. 
We verify that this is indeed the case by computing the ensemble-averaged end-to-end distance $\bm{R}^2$ [see Fig.~\ref{fig:diffusion}(a)]. 
We find no statistically significant change in the chain conformation upon the inclusion of our point-like obstacles for all chain lengths (with $d_T=4$) and tube diameters (with $N=200$). 
In Fig.~\ref{fig:diffusion}(b), we plot the monomer mean-square displacement (MSD) normalized by $t^{1/4}$ -- the characteristic power law that is a hallmark of reptation~\cite{Gennes1971}. 
In general, finite chain lengths can obscure observation of the expected reptation dynamics due to the fast relaxation of the chain-ends~\cite{Doi1986}. 
We therefore average the MSD over the interior $N_e$ beads and, in doing so, observe the emergence of a $t^{0.28}$ scaling in the MSD on timescales greater than the entanglement blob relaxation time and less than the Rouse time [see Fig.~\ref{fig:diffusion}(b)]. 
These findings support the idea that the forces exerted by the obstacles on the chain segments are purely entropic, allowing us to unambiguously explore the role of topological forces.

\subsection{Construction of the Primitive Chain}

While essential, construction of the primitive chain is in general highly nontrivial for multichain systems~\cite{Karayiannis2009}. 
However, for the obstacle model,  Edwards and Evans~\cite{EdwardsEvans1981} demonstrated that the primitive chain can be exactly defined; here, we briefly recapitulate their procedure and include a minor adaptation to ensure that the topological state of the chain is precisely captured.

\begin{figure}
	\centering
	\includegraphics[width=0.4\textwidth,keepaspectratio,clip]{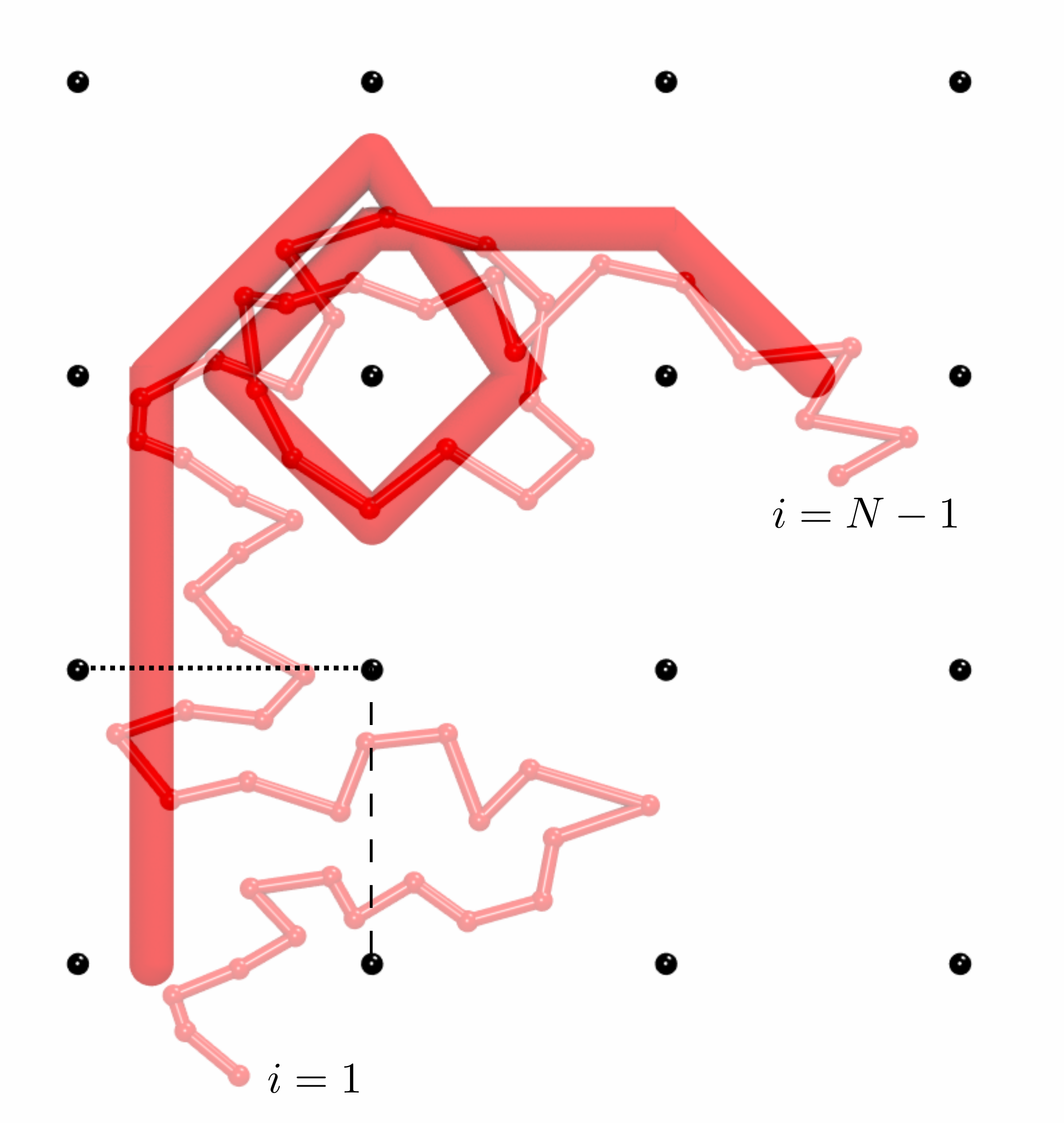}
	\caption{{\protect\small{The primitive chain (darker shading) constructed following the modified Edwards and Evans procedure. Overlapping primitive chain segments are offset for clarity.}}}
	\label{fig:pcanalysis}
\end{figure}

For a given chain configuration, we construct the primitive chain as we move along the segmental contour from one end of the chain ($i=1$) to the other ($i=N-1$) (see Fig.~\ref{fig:pcanalysis}), keeping track of each time a Kuhn segment crosses an edge  defined by two neighboring obstacles (see the dashed or dotted lines in Fig.~\ref{fig:pcanalysis}). 
If segment $i$ crosses an edge, the midpoint of the edge is part of the primitive path axis if every other segment $< i$ ($>i$) \textit{must} cross this edge to reach the position of the furthest chain-end $N-1$ ($1$). 
For a square lattice of obstacles, one can readily establish if this condition is met by keeping track of the number of times an edge is crossed, as demonstrated by Edwards and Evans~\cite{EdwardsEvans1981}. 
Consider the edge represented by the dashed line in Fig.~\ref{fig:pcanalysis}: the chain crosses this edge twice forming a loop which represents a lateral excursion from the primitive chain axis. 
In contrast, the edge denoted by the dotted line is crossed once, creating a new primitive chain segment. 
As we follow the chain contour we can distinguish excursions from the creation of new primitive chain segments by determining if a chain cross an edge an odd or even number of times. 
Counting these edge crossings must be done on a local-level e.~g., the total (or global) number of times an edge is crossed is not meaningful; rather, the number of times a chain crosses an edge before crossing the next edge is the relevant quantity. 
This consideration, which is not explicitly discussed by Edwards and Evans, is necessary in properly defining the primitive chain axis and is crucial for correctly capturing cases where the chain winds around a particular obstacle. 

\section{Results and Discussion}

\subsection{Quiescent Relaxation}

\begin{figure}
	\centering
	\includegraphics[width=0.48\textwidth,keepaspectratio,clip]{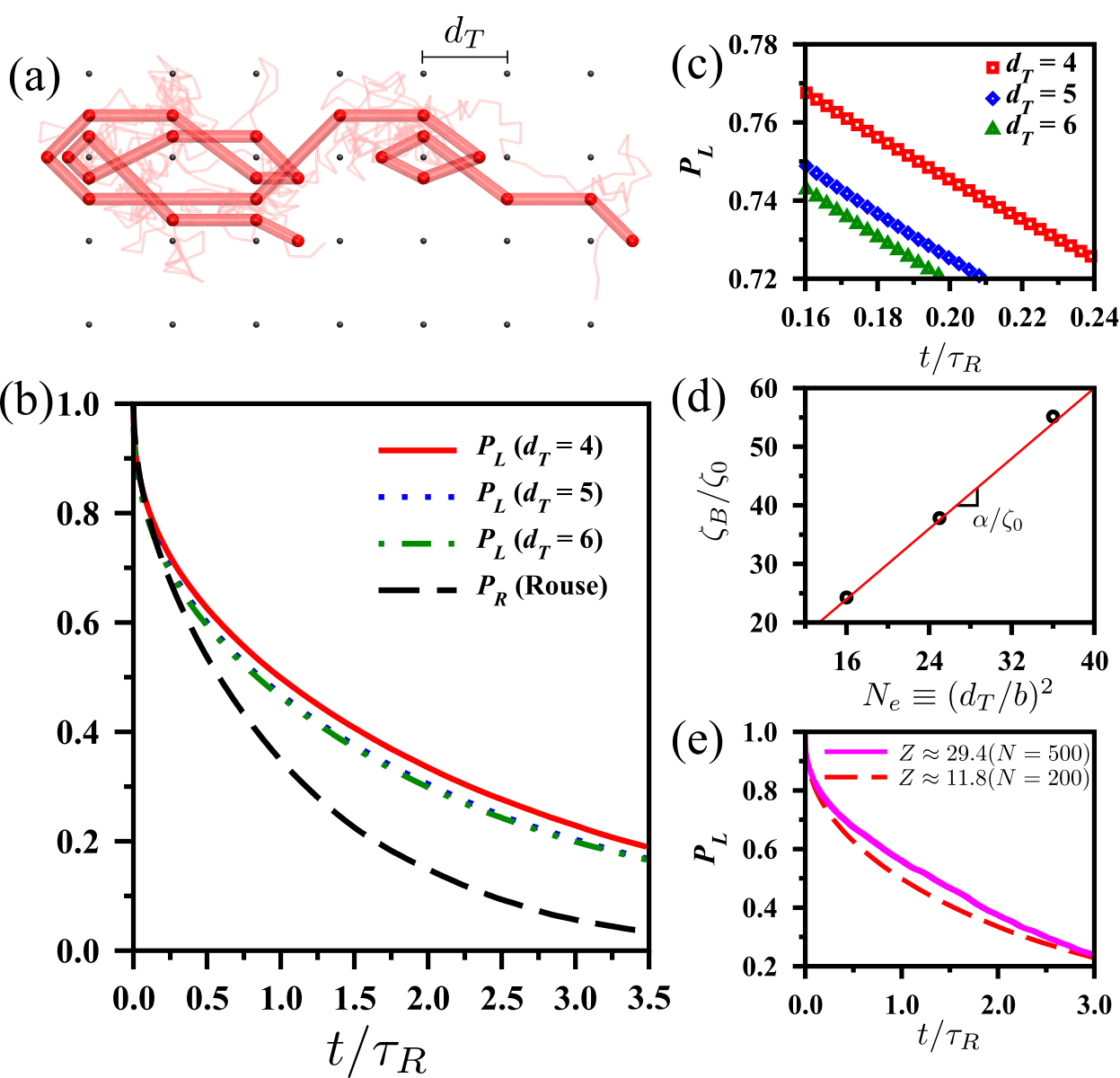}
	\caption{{\protect\small{(a) Snapshot of the primitive chain (dark shading) for a chain (light shading) for $N=500$ and $d_T = 4$. Overlapping primitive path segments are offset for clarity. (b) Simulation results for relaxation functions for $N=200$. (c) Early-time relaxation behavior. (d) $N_e$ dependence of blob-level friction. (e) $N$ dependence of $P_L$ with $d_T=4$. All data shown represent an average over $2000$ chains.}}}
	\label{fig:relaxation}
\end{figure}

Figure~\ref{fig:relaxation}(a) shows an instantaneous chain configuration and the resulting primitive chain using our analysis. 
We now explore the curvilinear dynamics in quiescence by examining the relaxation behavior of the primitive chain contour length ($L$) characterized by $\langle L(t)L(0) \rangle - \langle L \rangle^2$.  Rouse dynamics (for a continuous Rouse chain) for the curvilinear relaxation predicts~\cite{Doi1986}:
\begin{equation}
\label{eq:relaxeq}
P_L(t) \equiv \frac{\langle L(t)L(0) \rangle - \langle L \rangle^2}{\langle L^2 \rangle - \langle L \rangle^2} = \sum_{p=1 (\text{odd})}^{\infty}\frac{8}{\pi^2p^2}\exp \left [-\frac{p^2t}{\tau_R} \right],
\end{equation}
where $p$ represents the $p$th Rouse mode and $\tau_R = N^2b^2\zeta_0/2\pi^2kT$ is the (2D) Rouse time. 
$P_L(t)$ happens to be identical to the relaxation of the end-to-end distance $\bm{R}$ of a Rouse chain in the absence of the topological constraints (i.~e., $P_L=P_R \equiv \langle \bm{R}(t)\cdot\bm{R}(0) \rangle/Nb^2$). 
Computing $P_R$ for the free chain from simulation [see Fig.~\ref{fig:relaxation}(b)], we find the Rouse time extrapolated from the tail of $P_R$ to be the same as the analytical value. 

With unconstrained Rouse dynamics and the primitive path exactly defined, we can now quantitatively explore the curvilinear dynamics.  
Figure~\ref{fig:relaxation}(b) illustrates that for several values of the $d_T$ and $N=200$, contour length fluctuations, relax slower than the expected Rouse relaxation.
Fitting the tail of $P_L$ (between times of $2.0$ and $3.5\tau_R$) for each tube diameter, we find that the characteristic relaxation time $\tau$ is roughly $\tau \approx 2.5\tau_R$ \textit{for each tube diameter}. 
We interpret this slowed relaxation as evidence of a longitudinal topological force opposing curvilinear motion.

Interestingly, this topological force does not appear to \textit{qualitatively alter} the relaxation of the chain conformation from Rouse expectations [Eq.~\eqref{eq:relaxeq}].
Further, while the long-time relaxation is independent of $d_T$, there is a notable $d_T$ dependence at early times.
Careful examination of the early-time relaxation [see Fig.~\ref{fig:relaxation}(c)] reveals that the timescale at which this quantitative deviation from Rouse relaxation occurs correlates well with $d_T$ -- the length scale of the entanglement blobs. 
These observations (a quantitative slowdown occurring at the length scale of an entanglement blob) motivate the idea that the topological force can be coarse-grained into a blob-level ``friction" that slows down curvilinear relaxation of the blobs while preserving the underlying structure of the Rouse dynamics that gives rise to Eq.~\eqref{eq:relaxeq}. 
We introduce a phenomenological blob friction coefficient $\zeta_B$ such that the total curvilinear resistance felt by the chain (in addition to the Rouse friction $N\zeta_0$) due to the topological constraints is $Z \zeta_B$ where $Z \equiv N/N_e$ is the number of entanglement blobs and $N_e \equiv (d_T/b)^2$ (the average number of segments in a unit cell) is the entanglement molecular weight. 
The curvilinear relaxation time can then be expressed as:
\begin{equation}
    \label{eq:tau}
    \tau = \tau_R\left(1+\frac{Z\zeta_B}{N\zeta_0}\right). 
\end{equation}
The similar relaxation times for different $N_e$ indicates that the amount of friction $\zeta_B$ felt by the blob is linearly proportional to the blob size (thus $\zeta_BZ = {\rm const.}$), as shown in Fig.~\ref{fig:relaxation}(d) (the intercept of the plot is zero to within numerical accuracy). 
We identify the ratio $\zeta_B / N_e$ as an effective per-bead friction $\alpha$ that is invariant with the blob size ($\alpha \approx 1.5\zeta_0$) such that $\tau = \tau_R(1+\alpha/\zeta_0)$. 
Fixing $d_T$ and increasing the number of blobs (chain length) from $Z\approx11.8$ to $Z\approx29.4$ results in a relaxation response [see Fig.~\ref{fig:relaxation}(e)] that is consistent with a constant $\alpha$, in agreement with our expectations.

\subsection{Blob-level Topological Friction}

\begin{figure}
	\centering
	\includegraphics[width=0.48\textwidth,keepaspectratio, clip]{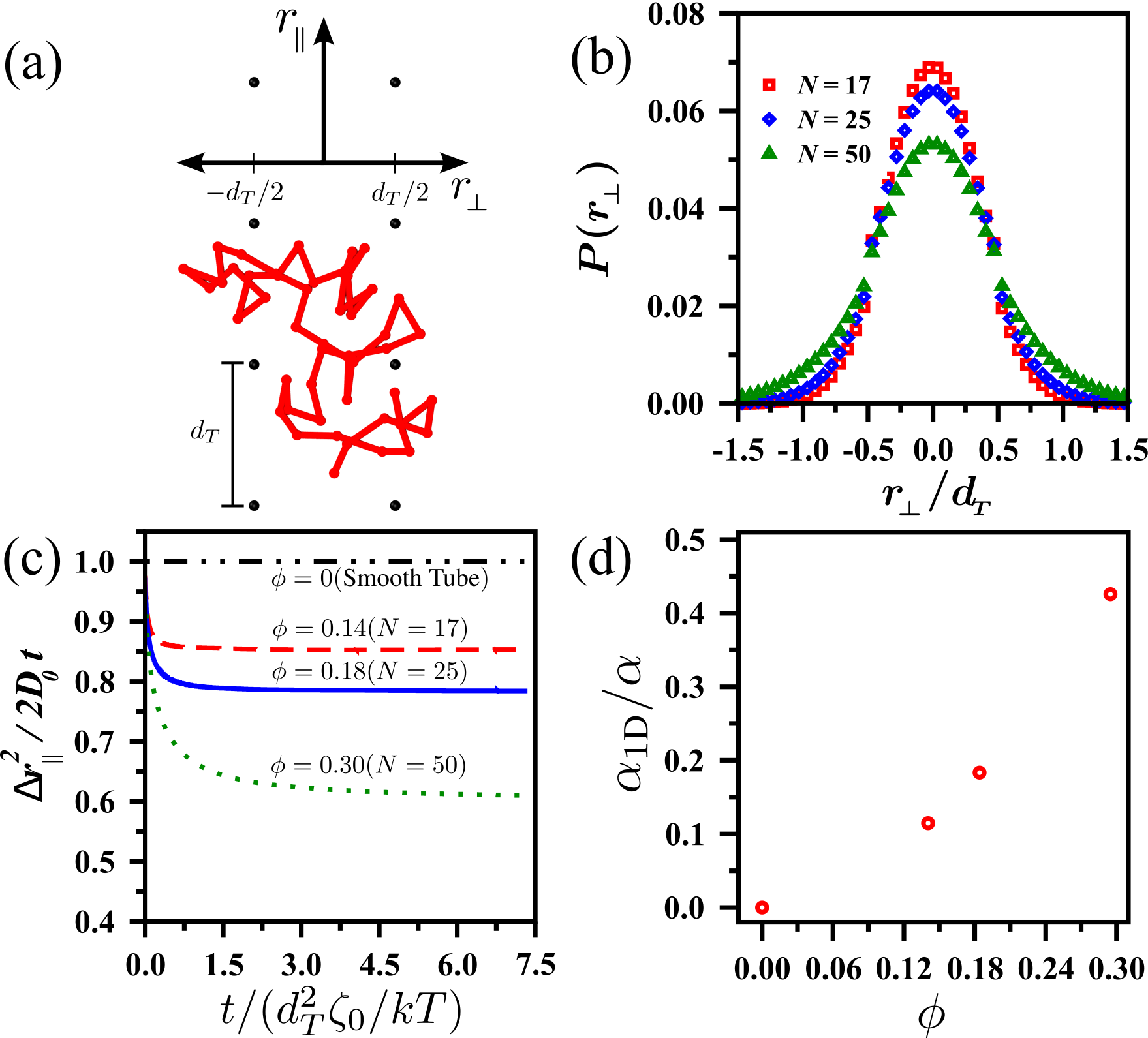}
	\caption{{\protect\small{(a) Simulation setup ($d_T=4$) for a blob with $N=50$. (b) Lateral segmental density distribution as a function of blob size. (c) Time dependence of the instantaneous longitudinal diffusion constant. (d) $\phi$ dependence of $\alpha_\text{1D}$.}}}
	\label{fig:friction}
\end{figure}

We speculate that this longitudinal topological force is related to segmental fluctuations outside of the tube (i.~e., tube ``leakiness"). 
Consider the system shown in Fig.~\ref{fig:friction}(a) -- a blob diffusing along a pseudo-1D tube with the ends of the blob constrained to remain inside tube (representing connectivity to a larger chain). 
As the blob moves along the tube (in the $r_{\parallel}$ direction), segments outside the tube (i.~e., beyond the obstacles in the transverse direction $r_{\perp}$) must overcome an entropic barrier to reenter the tube and bypass the impeding obstacles. 
The mechanical manifestation of this barrier is a topological force arising from the blob attempting to move longitudinally while some segments remain outside of the tube. 
This is directly related to the discrete nature of topological constraints, which introduces a spatially heterogeneous confining tube potential along the primitive chain from the perspective of a diffusing blob. 
Note that this entropic barrier is quite subtle and the size of the blob need not exceed $d_T$. 
Rather, this barrier is encountered by any blob having chain segments outside of the tube.

We explicitly test this idea by simulating the system depicted in Fig.~\ref{fig:friction}(a). 
For ideal chains, increasing $N$ naturally leads to an increase in the fraction of beads outside of the tube  $\phi = 1- \int_{-d_T/2}^{d_T/2}P(r_\perp)dr_\perp$ where $P(r_\perp)$ is the segmental position probability distribution in the transverse direction [Fig.~\ref{fig:friction}(b)]. 
We explore a large range of $\phi$ (by varying $N$) to capture the heterogeneity in blob size observed in our full 2D model [see Fig.~\ref{fig:relaxation}(a)]. 
For a smooth (achieved using hard walls) tube ($\phi=0$) the longitudinal mean-square-displacement of the chain center-of-mass $\Delta r_\parallel^2$ is diffusive at all times with a measured diffusivity of $D = D_0 = kT/N\zeta_0$ [see Fig.~\ref{fig:friction}(c)]. 
With increasing $\phi$, $\Delta r_\parallel^2$ becomes subdiffusive at early times as the blob motion localizes due to the topological force. 
To facilitate a comparison with our 2D model, we compute an effective per-bead excess friction constant $\alpha_\text{1D}$ by assuming the friction satisfies a Stokes-Einstein relation such that $\alpha_\text{1D} = kT/DN - \zeta_0$. 
We find $\alpha_\text{1D}$ to be a nonlinear function of $\phi$ [Fig.~\ref{fig:friction}(d)] -- the entropic barrier for tube reentry is strongly coupled to the number of beads that must reenter the tube. 
However, even for $\phi \approx 0.3$, $\alpha_\text{1D} < 0.5\alpha$ (where $\alpha$ is the topological friction coefficient measured in the full 2D [see Fig.~\ref{fig:relaxation}] model), suggesting that tube ``leakiness" and a single-blob perspective is insufficient in accounting for the observed friction in our full 2D model.

\subsection{Nonlinear Transient Microrheology}

{\em Theory for Transient Force Response.} The above discussions suggest that the winding/curved nature of the ``tube" (present in the full 2D model employed in Fig.~\ref{fig:relaxation} while absent in the model used in Fig.~\ref{fig:friction}) plays a significant role in generating the observed topological force.
Microrheology (in the full 2D model) offers a natural framework by which we can further probe the origins of this topological force. 
The force required to pull a chain-end at a fix velocity $v_p$ is equal and opposite to the sum of \textit{intermolecular}~\cite{Gennes1979} forces acting on the chain. 
In traditional tube physics, pulling the chain-end should result in the chain flowing along the tube with the only relevant intermolecular force being the curvilinear frictional force along the tube $N\zeta v_p$ with a per-bead friction of $\zeta_0$ in the absence of topological constraints.
If topological forces simply act to modify the effective per-bead friction, as suggested by the quiescent relaxation results (Fig.~\ref{fig:relaxation}), then the per-bead friction is $\zeta = \zeta_0 + \alpha$. 
We emphasize that this microrheology protocol (pulling a chain-end) should, in principle, solely probe curvilinear dynamics as the chain should freely flow along the tube in response to the pulling -- only needing to overcome the curvilinear resistance.

Consider a discrete Rouse chain consisting of $N$ beads. 
The beads are connected with Gaussian springs with a spring constant of $2kT/b^2$ such that the average spring length is simply $b$. 
We note that while stiffer bond springs were used in our simulation model, it can be readily shown that this only affects the short-time dynamics as one can always further coarse-grain the chain (group beads together to form ``super" beads) until the effective springs are indeed Gaussian. 
Classical reptation dynamics envisions that the effect of the obstacles is to restrict the Rouse motion of the chain to a tube-like path, reducing the Rouse equation of motion to 1D:
\begin{equation}
\label{eq:discreteRouse}
\zeta\dot{x}_n = -\frac{2kT}{b^2}\left(2x_n - x_{n+1} - x_{n-1}\right) + f_{B,n},
\end{equation}
where $x_n$, $\dot{x}_n$ and $f_{B,n}$ are the position (along the curvilinear coordinate of the tube), velocity and fluctuating Brownian force of the $n$th Rouse segment, respectively. 
We use a fixed velocity boundary condition $\dot{x}_0=-v_p$ on one end of the chain and introduce a tensile force boundary condition $x_{N+1}-x_{N}=\bar{L}/N$ on the other end to ensure an average equilibrium  contour length of $\bar{L}$ prior to pulling the chain.
($n=0$ and $n=N+1$ can be thought as ``phantom" beads affixed to the ends of the chain). The fixed velocity boundary condition can equivalently be expressed as a fixed position boundary condition with $x_0 = -v_pt$. 
The ensemble averaged force required to maintain this applied velocity is equal and opposite to the increase in the spring force on the pulled bead $-f(t) = (2kT/b^2)\langle x_0(t) - x_1(t) - x_0(0) + x_1(0)\rangle$. 

\begin{figure}
	\centering
	\includegraphics[width=0.48\textwidth,keepaspectratio,clip]{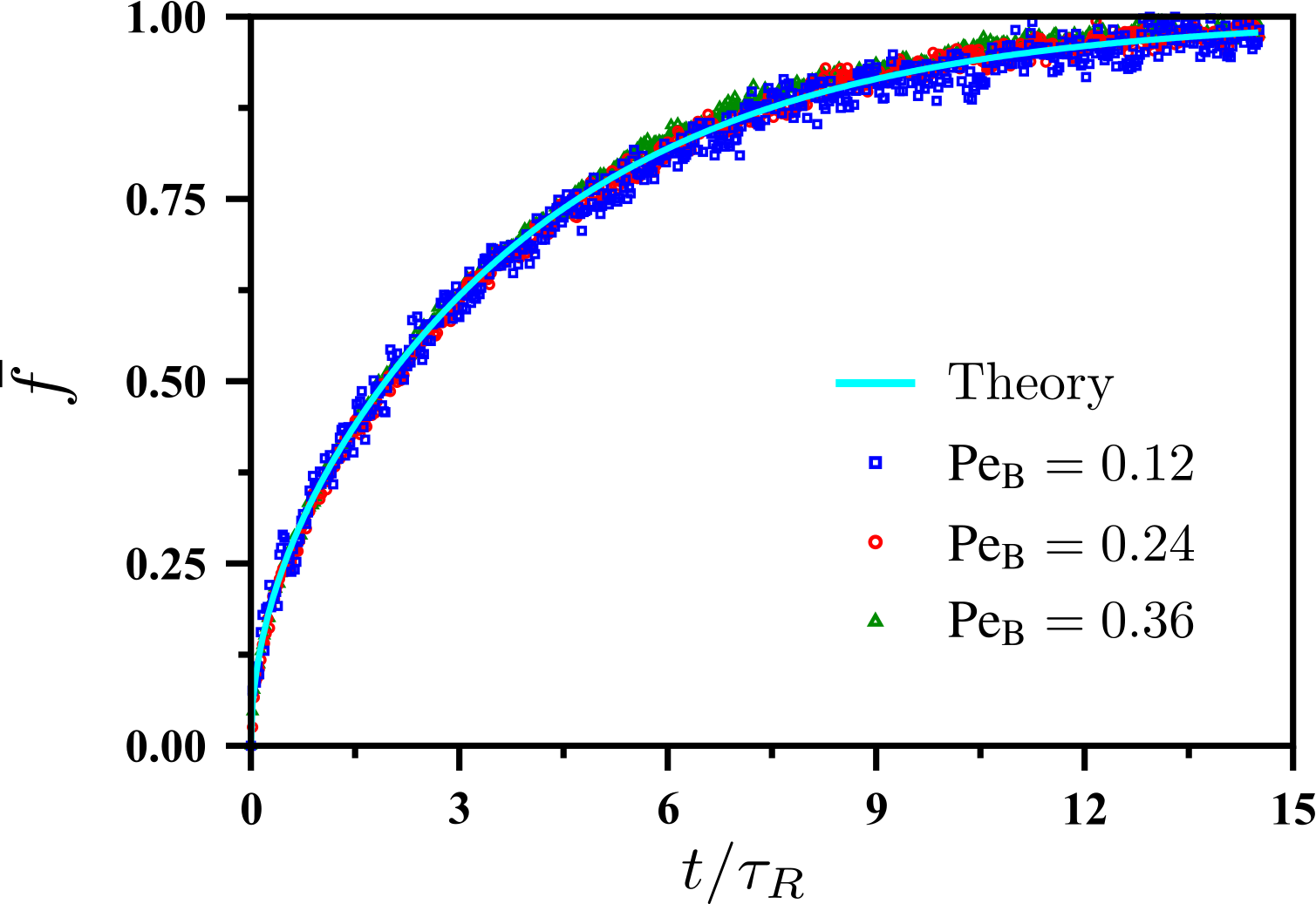}
	\caption{{\protect\small{Validation of Eq.~\eqref{eq:transientRouse} for Rouse chains ($N=200$) in the absence of obstacles. Without obstacles, the curvilinear relaxation time and steady-state force are simply $\tau_R$ and $N\zeta_0v_p$, respectively. All data is averaged over $6000$ independent chains.}}}
	\label{fig:rouse} 
\end{figure}

For long chains ($N\gg10$) we may approximate our discrete chain with a continuous chain parameterized with the Rouse curvilinear coordinate $n$~\cite{Doi1986}. 
Furthermore, as we only require the ensemble averaged positions we may average our equation of motion (eliminating the Brownian force) resulting in:
\begin{equation}
\label{eq:continuousRouse}
\zeta\frac{\partial \langle x_n \rangle}{\partial t} = \frac{2kT}{b^2}\frac{\partial^2 \langle x_n \rangle}{\partial n^2};
\end{equation}
\begin{equation}
\label{eq:continuousRouseBC1}
\langle x_0 \rangle = -v_pt; 
\end{equation}
\begin{equation}
\label{eq:continuousRouseBC2}
\frac{\partial \langle x_N \rangle}{\partial n} = \bar{L}/N.
\end{equation}
Note that the initial condition is simply $\langle x_n \rangle = n\bar{L}/N$. Equations~\eqref{eq:continuousRouse}-\eqref{eq:continuousRouseBC2} can be solved by assuming a solution of $\langle x_n(t) \rangle = -v_pt + n\bar{L}/N + \bar{x}(n,t)$.
Substituting this solution into the above equations results in an inhomogeneous equation for $\bar{x}(n,t)$ with homogeneous boundary conditions. 
$\bar{x}(n,t)$ can be expressed with normal modes $\bar{x}(n,t) = \sum_{p=1\text{(odd)}}^{ \infty }x_p(t) \sin(\lambda_p n)$ where $\lambda_p = p\pi/2N$ and $p$ is an odd integer. 
Solving for the Fourier coefficients $x_p(t)$ we find:
\begin{equation}
\label{eq:fourier}
-x_p(t) = \frac{\zeta v_pb^2}{N kT \lambda_p^3}\left(1-e^{-p^2t/4\tau}\right),
\end{equation}
where $\tau=N^2b^2\zeta/2\pi^2kT$ is the curvilinear relaxation time for a 2D chain. 
We determine the transient force response by substituting our full solution $\langle x_n(t) \rangle = -v_pt + n\bar{L}/N + \sum_{p=1\text{(odd)}}^{ \infty }x_p \sin(\lambda_p n)$ into:
\begin{equation}
\label{eq:force}
-f(t) = \frac{2kT}{b^2}\left(\frac{\partial \langle x_0(t) \rangle}{\partial n} - \frac{\partial \langle x_0(0) \rangle}{\partial n}\right),
\end{equation}
which results in:
\begin{equation}
\label{eq:transientRouse}
-\frac{{f}(t/\tau)}{N\zeta v_p} \equiv \bar{f}(t/\tau) = 1-\sum_{p=1 (\text{odd})}^{\infty}\frac{8}{\pi^2p^2}\exp \left [-\frac{p^2t}{4\tau} \right].
\end{equation}
Note that the tube diameter or contour length are absent in Eq.~\eqref{eq:transientRouse} -- the only resistance to motion is the curvilinear friction along the tube that resists the Rouse motion. 
Indeed, one can readily show that Eq.~\eqref{eq:transientRouse} is rigorously exact for unconstrained Rouse chains as well. 
We therefore verify Eq.~\eqref{eq:transientRouse} with Brownian dynamics simulations of Rouse chains in the absence of obstacles and find perfect agreement as shown in Fig.~\ref{fig:rouse}. 
In the absence of obstacles, the per-bead friction coefficient is simply $\zeta = \zeta_0$ and thus the curvilinear relaxation time is the Rouse time $\tau_R$.   

The required pulling force increases with time as larger sections of the chain are being dragged while maintaining the speed of the chain-end. 
Interestingly, Eq.~\eqref{eq:transientRouse} predicts that the force-response for all pulling speeds will fall along a master-curve and, further, will reach a steady-state with a characteristic timescale that is four times the curvilinear relaxation time (i.~e., 4$\tau \approx 10 \tau_R$ with $\zeta = \zeta_0 + \alpha \approx 2.5 \zeta_0$ ). 
Evaluating the applicability of Eq.~\eqref{eq:transientRouse} to our simulation data will allow us to asses the validity of representing the topological force as a local friction that is independent of chain conformation.

\begin{figure*}
	\centering
	\includegraphics[width=0.96\textwidth,keepaspectratio,clip]{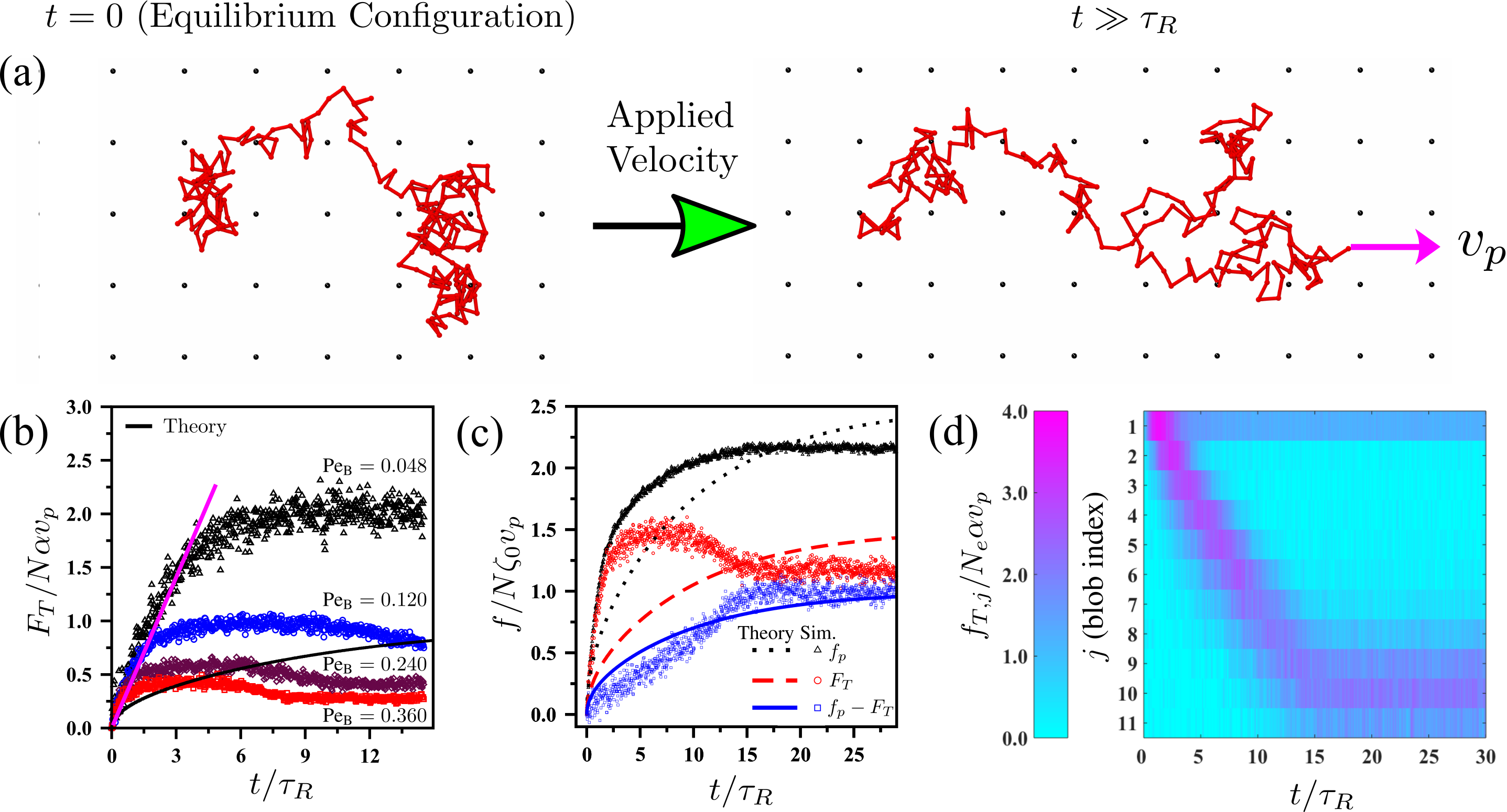}
	\caption{{\protect\small{(a) Microrheology setup ($d_T=4$) illustrating a typical initial (top) and final (bottom) configuration of a chain with $N=200$. (b) $\text{Pe}_B$ dependence of the transient evolution of $F_T$. (c) Breakdown and theoretical comparison of force response for $\text{Pe}_B=0.12$. (d) Blob-level transient $f_{T,j}$ response for $\text{Pe}_B=0.12$. All data is averaged over $6000$ chains. }}}
	\label{fig:microrheology}
\end{figure*}

{\em Simulation Results.} We focus on velocities in the nonlinear regime with respect to orientational relaxation (i.~e., $v_p\gg R/\tau_D$ where $R$ and $\tau_D$ are, respectively, the end-to-end distance and reptation time of the chain). 
The total ($f_p$) force required to pull the chain is the sum of a topological $F_T$ (the total amount of force exerted by the obstacles on the chain) and non-topological $f_p-F_T$ force. 
We compare our measured forces $f_p$ and $F_T$ with the theoretical expectations of $f_p = N\zeta v_p\bar{f}(t/\tau)$, $F_T = N\alpha v_p\bar{f}(t/\tau)$ and $f_p-F_T = N\zeta_0v_p\bar{f}(t/\tau)$. 
We pull the chain-end (which is placed in the center of a unit cell) along the lattice axis such that the chain is fully aligned along the pulling axis at steady-state [see Fig.~\ref{fig:microrheology}(a)].

Figure~\ref{fig:microrheology}(b) shows the evolution of $F_T$ for various rates, expressed in terms of a ``blob" P\'{e}clet number defined as $\text{Pe}_B \equiv v_pd_T/D_B$ where $D_B = kT/N_e\zeta_0$. 
At the lowest pulling rate, $F_T$  rises linearly to twice the anticipated steady-state force of $N\alpha v_p$, suggesting a nonlinear force-velocity relationship. 
With increasing $\text{Pe}_B$, the steady-state $F_T$ increases sublinearly (i.~.e., $F_T/v_p$ decreases with increasing $v_p$, a ``force-thinning" response) and eventually subceeds $N\alpha v_p$.  
At the higher rates, $F_T$ exhibits an overshoot before reaching steady-state while the non-topological drag force ($f_p - F_T$) continues to increase linearly to its terminal value ($N\zeta_0v_p$) as shown in Fig.~\ref{fig:microrheology}(c).  
All these observations are in stark contrast to the predicted Rouse response [Eq.~\eqref{eq:transientRouse}].

It proves revealing to explore the evolution of the topological force at the blob-level. 
We divide the chain into $Z$ blobs with $N_e$ beads per blob and compute the total topological force (in the pulling direction) $f_{T,j}$ acting on the $j$th blob (where $j=1$ corresponds to the blob directly connected to the pulled bead) by summing over the obstacle force on each bead within the blob. 
Remarkably, we find [see Fig.~\ref{fig:microrheology}(d)] that $f_{T,j=1}$ on the primary ($j=1$) blob undergoes a striking overshoot with a maximum that exceeds the expected steady-state value of $N_e\alpha v_p$ by nearly a factor of four. 
As the force on the primary blob begins to decrease, the force on the next blob begins a similar trajectory. 
The process repeats for each blob resulting in a curvilinear cascade of strong overshoots in the topological force. 
Interestingly, the last few blobs experience little overshoot and are able to maintain a relatively high force. 

\begin{figure}
	\centering
	\includegraphics[width=0.48\textwidth,keepaspectratio,clip]{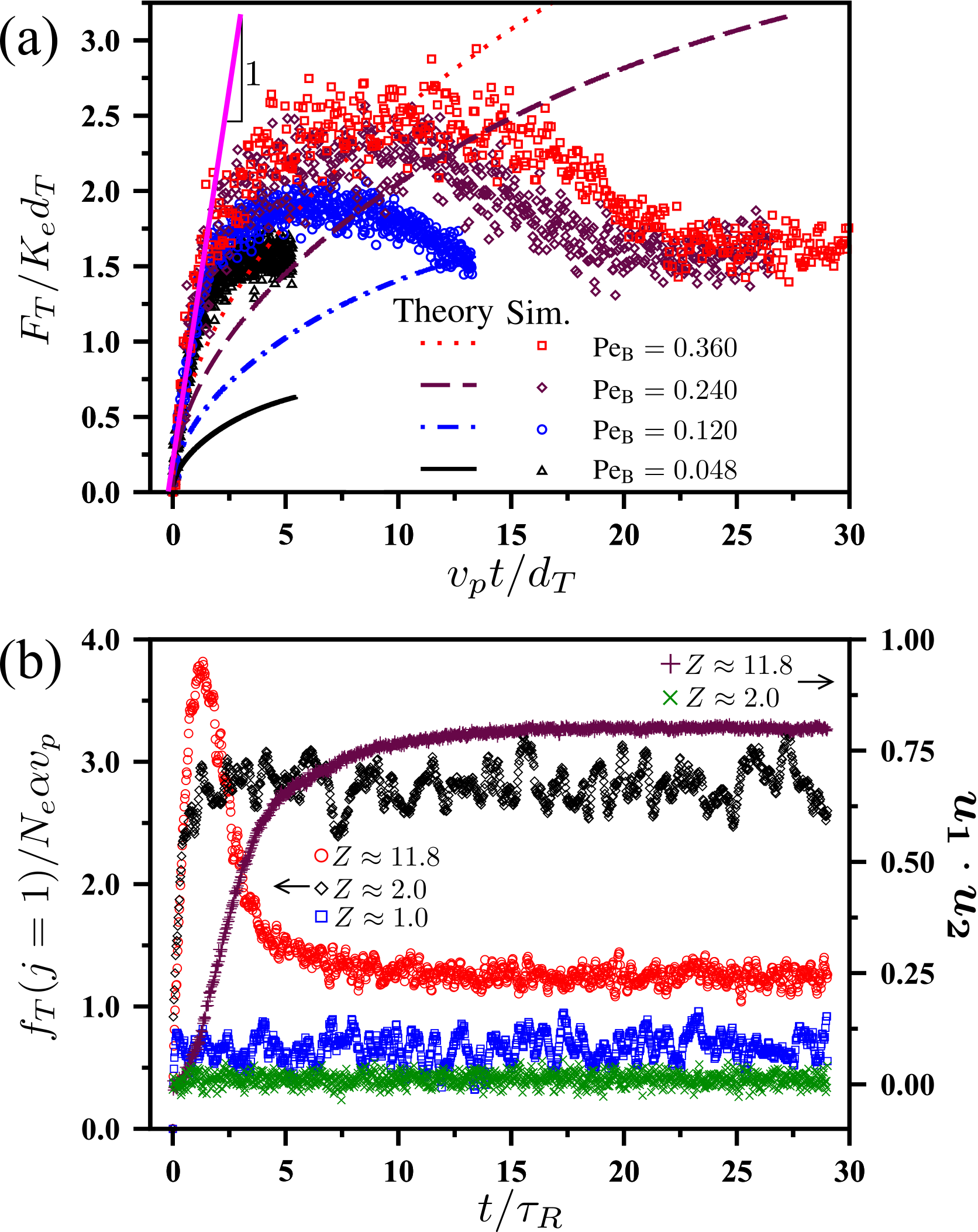}
	\caption{{\protect\small{Startup response for $d_T=4$ of: (a) $F_T$ as a function of displacement and $\text{Pe}_B$; and (b) $f_T$ of the primary blob ($j=1$) and $\bm{u_1\cdot u_2}$ as a function of $Z$ ($N$) with $\text{Pe}_B = 0.12$.}}}
	\label{fig:elasticresponse}
\end{figure}

The observed overshoot cascade coupled with the initial linear rise of the topological force for all $\text{Pe}_B$ is highly suggestive of a blob-level elastic response -- with the topological forces balancing intramolecular elastic forces. 
Indeed, in plotting the force $F_T$ as a function of the net displacement of the pulled bead $v_pt$, we find an initial linear slope that is independent of the pulling rate as shown in Fig.~\ref{fig:elasticresponse}(a). 
This is in marked contrast to the Rouse response [Eq.~\eqref{eq:transientRouse}] wherein the transient force depends only on the time $t$ scaled by the curvilinear relaxation time $\tau$. 
The initial linear slope (the effective spring constant $K_e$) corresponds to a blob of size $2kT/K_eb^2 \approx 1.5N_e$, in agreement with a blob-level response. 
Interestingly, the displacement (or force) at which the elastic response begins to fail is not a universal value, but rather, increases (sublinearly) with the pulling rate.

Further insight into the nature of this topology-induced elasticity can be found in comparing the total topological force on the primary blobs of chains of various size $N$ at a fixed $\text{Pe}_B$ as shown in Fig.~\ref{fig:elasticresponse}(b). 
A single blob ($N=N_e$) exhibits an entirely dissipative response at all times [blue squares in Fig.~\ref{fig:elasticresponse}(b)], highlighting the nonlocal and multiblob origins of the topological force. 
Interestingly, two blobs ($N=2N_e$) is sufficient to generate a linear rise in the $f_{T,j=1}$ [black diamonds in Fig.~\ref{fig:elasticresponse}(b)] with an identical slope to our full chain $N\approx 11.8N_e$ [red circles in Fig.~\ref{fig:elasticresponse}(b)]. 
However, the elastic response of the shorter chain ceases earlier than that of the full chain and, rather than overshoot, plateaus. 
Further, the steady-state $f_{T,j=1}$ for the shorter chain is nearly three times \textit{larger} than that of the full chain. 
This contrast in steady-state forces correlates well with the relative orientation of the first two blobs $\bm{u_1\cdot u_2}$ (where $\bm{u_j}$ is the unit vector of the blob end-to-end distance). 
After the full chain $f_{T,j=1}$ reaches a maximum, $\bm{u_1\cdot u_2}$ rises monotonically before saturating, coinciding with the decline and plateau of $f_{T,j=1}$ [maroon data in Fig.~\ref{fig:elasticresponse}(b)]. 
In contrast, the second blob of the shorter chain is less constrained and free to relax, resulting in $\bm{u_1\cdot u_2}\approx 0$ at all times [green data in Fig.~\ref{fig:elasticresponse}(b)]. 
This coupling of the topological force to the orientation is consistent with our earlier speculation that the tube curvature plays an important role in generating topological forces and appears to support the idea that the topological force generating the elastic response is coupled to orientational relaxation, which was recently invoked in attempting to theoretically reconcile experimental observations in the nonlinear rheology of entangled polymers~\cite{Schweizer2018}. 

\section{Conclusions and Outlook} 

Our work demonstrates that even for the simplest model system for topologically constrained chains, topological forces can have a profound impact on chain dynamics that goes {\em beyond} restricting chain motion to the primitive path. 
In quiescence, topological forces projected onto the primitive path are found to increase the curvilinear relaxation time. While these forces strongly resemble a curvilinear topological ``friction" that can be thought of as a shift-factor to the traditional Rouse friction, the origins of this force appear to be an inherently multiblob nonlocal (albeit on a length scale below the size of the chain) effect that depends on the winding nature of the tube.  
Microrheology reveals a pronounced coupling of topological forces with chain conformation, manifesting in a marked deviation of the (transient) force-velocity relation from the expected Rouse response. 
These nonequilibrium topological forces are a highly nonlinear function of deformation, stalling blobs from freely flowing, and inducing an elastic response. 

Our observations suggest that the far-from-equilibrium dynamics involves new physical considerations that cannot be obtained simply from extrapolation of the quiescent-state physics.  
In particular, we believe the discreteness of the topological interactions to be essential in the nonlinear dynamics of topologically constrained polymers, such as entangled polymer melts and solutions.  
Isolating these effects in full molecular dynamics simulations of entangled polymers~\cite{Everaers2004, Cao2012, Xu2018, Ruan2021}, while an exciting and important challenge, is complicated by the ambiguity in defining topological forces. 
It may prove useful to next explore a model at an intermediate level of complexity in comparison to the system employed in this work and full molecular dynamics simulations. 
Models that explicitly retain discrete topological interactions while more closely resembling entangled polymers, such as slip link simulations~\cite{Masubuchi2001,Chappa2012,Ramirez-Hernandez2015}, may provide further insight into the nature and roles of these topological interactions in far-from-equilibrium entangled polymer liquids. 

\begin{acknowledgments} 
A.K.O. and Z.-G.W. thank Umi Yamamoto and Kevin Shen for helpful discussions. 
Y.L. and L.A. acknowledge support by the Natural Science Foundation of China under Grant Nos. 21790340 and 21120102037.
\end{acknowledgments}

\end{document}